\documentclass[letterpaper,conference,final]{IEEEtran}
\IEEEoverridecommandlockouts
\usepackage[utf8]{inputenc}
\usepackage[T1]{fontenc}
\usepackage[english]{babel}
\usepackage{ntheorem}
\usepackage{amsmath,amssymb,amstext}
\usepackage{epsfig,psfrag}
\usepackage{graphics,graphicx} 
\usepackage{color}
\usepackage{dsfont}
\usepackage{url}
\usepackage{alltt}
\usepackage{stmaryrd}
\usepackage{color}
\usepackage[font=footnotesize]{subfig}
\usepackage{array,algorithm2e,algorithmic}

\usepackage[width=18.2cm,height=25.7cm]{geometry}
\usepackage{epsfig}
\usepackage{flushend}

\newtheorem{definition}{Definition}

\newtheorem{theorem}{Theorem}

\newtheorem{lemma}{Lemma}

\SetKwInput{KwData}{Parameters}

\newcommand{\Nats}[0]{\ensuremath{\mathds{N}}}

\newcommand{\R}[0]{\ensuremath{\mathds{R}}}
\newcommand{\Bool}[0]{\ensuremath{\mathds{B}}}

\renewcommand{\le}{\leqslant}
\renewcommand{\ge}{\geqslant}

\begin{document}

\title{Quality Analysis of a Chaotic Proven Keyed Hash Function}

\author{Jacques M. Bahi, Jean-Fran\c{c}ois Couchot, and Christophe Guyeux\\
University of Franche-Comté, FEMTO-ST Institute\\
Belfort, France\\
Email: \{jacques.bahi, jean-francois.couchot, christophe.guyeux\}@univ-fcomte.fr}

\maketitle

\begin{abstract}
Hash functions are cryptographic tools, which are notably involved
in integrity checking and password storage.
They are of primary importance to improve the security
of exchanges through the Internet.
However, as security flaws have been recently identified in the current standard in this domain, new ways to hash digital data must be investigated.
In this document an original keyed hash function is evaluated.
It is based on asynchronous iterations leading to functions 
that have been proven to be chaotic.
It thus possesses various topological properties as uniformity and sensibility to its initial condition.
These properties make our hash function satisfies established security 
requirements in this field.
This claim is qualitatively proven and experimentally verified in this research work, among other things by realizing a large number of simulations.

\end{abstract}


\section{Introduction}\label{sec:intro}
The security and the
 privacy of data exchanged through the Internet are guaranteed
by protocols that make an adequate use of a few cryptographic tools as
secure pseudorandom number generators or hash functions. Hash functions
are applications that map words of any lengths to words of fixed lengths
(often 256 or 512 bits). These hash functions allow, for instance,
 to  store passwords  in a  secure manner or  to check  whether a
download has occurred  without any error.  
They be designed to depend from a given parameter, 
called a key. According to their field of application, the requirements
an hash function has to satisfy can change. They need at least to be
very fast, so that the diffusion of the digest into the set of hash
values occurs (whatever the bias into the inputted message), and so
that a link between a message and its digest is impossible to 
establish in practice (confusion). The possibility to use a key or to
distribute the computation on numerous threads must often be offered 
in several applications.
Finally, in the computer security field, stringent complexity properties
have to be proven, namely the collision, preimage, and 
second-preimage resistances, the unpredictability, and the 
pseudo-randomness properties. Each of the latter one have a rigorous 
formulation in terms of polynomial indistinguishability.

Several hash functions have been proposed as candidates to be standards
in computer science. Such standards are designed by the scientific community
and selected, after peer studies, by administrations as the NIST one
(National Institute for Standards and Technologies of the US government).
SHA-1 is  probably the most widely  used  hash function.   
It  is present  in  a  large panel  of
security applications and protocols through the Internet.  

However, in 2004, MD5 and SHA-0 have been broken. 
An attack over SHA-1 has been achieved with only $2^{69}$ operations (CRYPTO-2005), that is, $2,000$ times faster than a brute force attack (that requires $2^{80}$ operations).
Even if $2^{69}$ operations still remain impossible to realize on common computers, such a result, based on a previous attack on SHA-0, is a very important one:
as the SHA-2 variants are algorithmically  close to SHA-1 and 
eventually produce  message  digests on  principles similar 
to  the MD4  and MD5  message digest algorithms, a new hash standard based on original approaches is
then eagerly  awaited.  
This is why a SHA-3 contest has been launched these last few years, to find
a new, more secure standard for hash functions.
So new original hash functions, or improvements for existing ones, must be found.

In this context,  we have proposed  a new hash
function  in~\cite{bcg11:ip,guyeux09}.   
Being designed by using discrete dynamical systems, and taking benefits
from various established topological properties, this new family of
hash functions is thus based on a completely different approach. Among
other things, in our proposal, an ingredient of chaos is added to existing
hash functions, in order to reinforce their properties.
Introducing chaos into the design of hash functions
has been already 
addressed in~\cite{Wang2003,Xiao20094346,Xiao20092288,Xiao20102254}.
These methods usually
transform the initial message into its padded fixed length version
and then translate it into a real number.
Next, with a chosen chaotic map (some chaotic  functions of
real variables like logistic, tent,  or Arnold's cat maps, for instance~\cite{web:listOfChaoticMaps}),
methods set the initial algorithm parameters 
according to the secret key and start iterations.
Methods are then left to extract some bits from the iterations results
and to juxtapose them to get the hash value.
It is  then supposed
that  the final  hash  function  preserves the  properties of chaos.
However, the idea of chaotic hash functions has been
controversially discussed in the community~\cite{Zhou1997429,Guo20093201}.
Moreover,  even if  these
algorithms are themselves proven  to be chaotic, their implementations
on finite machines can result  into the loss of chaos property.  Among other
things, the main  reason is that chaotic functions  (embedded in these
researches)  only manipulate  real numbers,  which do  not exist  in a
computer.   In~\cite{guyeux09}, the  hash  function we  have
proposed does not simply integrate chaotic maps into algorithms hoping
that the  result remains chaotic;  we have conceived an  algorithm and
have mathematically proven that it  is chaotic.  To do both, our theory
and our implementation are based on finite integer domains 
and finite states iterations, where only one randomly chosen element
is modified at each step. This iteration mode is further referred to as 
asynchronous mode.

These studies  lead to the  conclusion that the  chaos of asynchronous 
iterations is very intense~\cite{GuyeuxThese10}.  
As this mode only manipulates  binary digits or
integers,  we have  shown  that   truly
chaotic  computer programs can be produced. They can thus be applied to
pseudorandom       number
generators~\cite{bcgw11:ip}  and   to  a complete class of 
information  hiding schemes~\cite{bcg11:ij}, for instance.    
In  this  paper,  the  complete chaotic behavior of asynchronous iterations
is capitalized to produce a truly chaotic keyed hash function.

This research work is an improvement of a previous article accepted at
the Third International Conference on Evolving Internet, INTERNET11
(June 19-24, 2011, Luxembourg)~\cite{bcg11:ip}. Compared to this
research work, the proposed hash function (Section~\ref{sec:algo})
has been completely rethought. It
appears now more as a post-treatment on existing hash functions, to 
improve their security (Sections~\ref{sec:algo}, \ref{sec:eval}), than as 
a hash function designed from scratch. Moreover, the second-preimage resistance
has been proven in Section~\ref{preimageProof} and the 
strict avalanche criterion has been statistically studied 
(Section~\ref{sub:sac}). All these improvements
lead to obviously better scores for the proposed hash functions, when
experimentally evaluating  its security.

The remainder of this research work is organized in the following way.
In   Section~\ref{sec:chaos},   basic   notions  concerning   asynchronous
iterations and Devaney's chaos  are recalled.  Our keyed hash function
is  presented in  Section~\ref{sec:algo}.
Performance analyses  are presented in  the next two sections:  in the
first  one a  qualitative  evaluation of  this  function is  outlined,
whereas  in  the second  one  it  is  evaluated experimentally.   This
research work ends by a  conclusion section, in which our contribution is
summarized and intended future work is mentioned.

\section{Background Section}\label{sec:chaos}
In this section,  we first give definitions 
of Secure Keyed One-Way Hash Functions and of  
the Strict Avalanche Criterion (SAC), which is a property 
that such a function has to verify.
Next we  give  some recalls  on Boolean discrete dynamical systems and 
link them with topological chaos.
Finally,  we establish relations between  the algorithm properties inherited
from topological results and the requirements of Secure Keyed One-Way Hash
Function.

\subsection{Secure Keyed One-Way Hash Function}

\begin{definition}[Secure Keyed One-Way Hash Function~\cite{BSP96}]
Let $\Gamma$ and $\Sigma$ be two alphabets,   
let $k \in K$ be a key in a given key space,
let $l$ be a natural number, which is the length of the output message,
and let $h : K \times  \Gamma^{+} \rightarrow \Sigma^{l}$ be a function that associates 
a message in $\Sigma^{l}$ for each pair of key, word in  
$K \times  \Gamma^{+}$.
The set of all functions $h$ is partitioned into classes
of functions $\{h_k : k \in K \}$
indexed by a key $k$ and such that 
$h_{k}: \Gamma^{+} \rightarrow \Sigma^{l}$ is defined by
$h_{k}(m) = h(k,m)$,  \textit{i.e.}, $h_{k}$ generates a message digest of length $l$.

A class $\{h_k : k \in K \}$ is a \emph{Secure Keyed One-Way Hash Function}  
if it satisfies the following properties:
\begin{enumerate}
\item the function $h_k$ is \emph{keyed one-way}. That is,
  \begin{enumerate}
  \item Given $k$ and $m$, it is easy to compute $h_k(m)$.
  \item Without the full knowledge of $k$, it is 
    \begin{itemize} 
      \item difficult to find $m$ when $h_k(m)$ is given;
        this property is referred to as \emph{preimage resistance};
      \item difficult to find $h_k(m)$ when only $m$ is given. 
      \end{itemize}
    \end{enumerate}
\item The function $h_k$ is the \emph{keyed collision resistant}, that is, 
  without the knowledge of $k$ it is difficult to find two distinct messages
  $m$ and $m'$ s.t. $h_k(m)= h_k(m')$.
  A weaker version of this property is the \emph{second preimage resistance},
  which is established if for a given $m$ it is 
  difficult to find another message $m'$, $m \neq m'$, such that 
  $h_k(m)= h_k(m')$.

\item Images of function $h_k$ have to be uniformly distributed in 
$\Sigma^{l}$ in order to counter statistical attacks. 
\item Length $l$ of the produced image has to be larger than $128$ bits 
 in order to counter birthday attacks~\cite{DBLP:conf/crypto/Coppersmith85}.
\item Key space size has to be sufficiently large 
in order to counter exhaustive key search.
\end{enumerate}
\end{definition}

Finally, hash functions have to verify the \emph{strict avalanche  criterion} 
defined as follows:
\begin{definition}[Strict Avalanche Criterion~\cite{Webster:1985:DS:646751.704578}]
Let $x$ and $\overline{x}^i$, two $n$-bit, binary vectors, 
such that $x$ and $\overline{x}^i$ differ  
only in bit $i$, $1 \le i  \le n$. Let $f$ be the cryptographic transformation 
(hash function applied on vector of bits for instance). 
Let   $\oplus$ be the exclusive or operator. 
The $f$ function  meets the \emph{strict avalanche criterion} 
if and only if the following property is established;
$$
\begin{array}{l}
\forall n. \, \forall i,j.\,  1 \le i \le n \land 1 \le j \le m  \Rightarrow
\\ 
\qquad \qquad  P \left( \big(f(x) \oplus f(\overline{x}^i)\big)_j = 1\right) = 1/2
\end{array}
$$
\label{def:sac}
\end{definition}
This means that for any length message, each bit of the digest is independent 
of modifying one bit in the original message.  
In other words, a difference  of  one  bit between
two  given  medias  has to  lead  to completely different digests.

\subsection{Boolean Discrete Dynamical Systems}

Let us first discuss the  domain of iterated functions.  As far as
we know, no result rules that  the chaotic behavior of a function that
has  been   theoretically  proven  on   $\R$  remains  valid   on  the
floating-point numbers, which is  the implementation domain.  Thus, to
avoid the loss of chaos this research work presents an alternative, namely
to iterate  Boolean maps: results  that are theoretically  obtained in
that domain are preserved during implementations.
This section recalls facts concerning Boolean discrete-time 
dynamical Systems (BS) that are sufficient to understand the background
of our approach.

Let us denote by $\llbracket a ; b \rrbracket$ the interval of integers:
$\{a, a+1, \hdots, b\}$, where $a \leqslant b$. 
Let $n$ be a positive integer. A Boolean discrete-time 
system is a discrete dynamical
system defined from a {\emph{Boolean map}}
$f:\Bool^n\to\Bool^n$ s.t. 
\[
  x=(x_1,\dots,x_n)\mapsto f(x)=(f_1(x),\dots,f_n(x)),
\]
{\emph{and an iteration scheme}}: parallel, asynchronous\ldots 
With the \emph{parallel} iteration scheme, 
the dynamics of the system are described by $x^{t+1}=f(x^t)$
where $x^0 \in \Bool^n$.
Let thus $F_f: \llbracket1;n\rrbracket\times \Bool^{n}$ to $\Bool^n$ 
be defined by
\[
F_f(i,x)=(x_1,\dots,x_{i-1},f_i(x),x_{i+1},\dots,x_n),
\]
with the \emph{asynchronous} scheme,
the dynamics of the system are described by $x^{t+1}=F_f(s^t,x^t)$
where $x^0\in\Bool^n$ and $s$ is a {\emph{strategy}}, \textit{i.e.}, a sequence 
in $\llbracket1;n\rrbracket^\Nats$.
Notice that this scheme only modifies one element at each iteration.

Let $G_f$ be the map from $\mathcal{X}= \llbracket1;n\rrbracket^\Nats\times\Bool^n$ to 
itself s.t.
\[
G_f(s,x)=(\sigma(s),F_f(s^0,x)),
\] 
where $\sigma(s)^t=s^{t+1}$ for all $t$ in $\Nats$. 
Notice that the parallel iteration of $G_f$ from an initial point
$X^0=(s,x^0)$ describes the ``same dynamics'' as the asynchronous
iteration of $f$ induced by the initial point $x^0$ and the strategy
$s$.

The state-vector  $x^{t}= (x_1^{t},  \ldots, x_{n}^{t})  \in  \mathds{B}^n$
of the system at discrete time $t$ (also  said at {\em iteration} $t$) is 
further denoted as the  {\em configuration} of the system at time $t$.

In what follows,  the dynamics of  the system is particularized  with the
negation function $\neg :  \Bool^n \rightarrow \mathds{B}^n$ such that
$\neg(x)    =   (\overline{x_i},   \ldots,    \overline{x_n})$   where
$\overline{x_i}$ is the negation of $x_i$.
We thus have the function $F_{\neg}$ that is
defined by:
$$\begin{array}{ccl}
F_{\neg}:  \llbracket1;n\rrbracket\times \mathds{B}^{n}
& \rightarrow  & 
\mathds{B}^{n} \\  
F_{\neg}(s,x)_j & =  & 

\left\{
\begin{array}{l}
\overline{x_j} \textrm{ if } j= s \\ 
x_{j} \textrm{ otherwise.} \enspace 
\end{array}
\right. 
\end{array}$$

\noindent With such a notation, configurations are defined for times 
$t=0,1,2,\ldots$ by:
\begin{equation}\label{eq:sync}   \left\{\begin{array}{l}   x^{0}\in
\mathds{B}^{n} \textrm{ and}\\
 x^{t+1}= F_{\neg}(S^t,x^{t}) \enspace .
\end{array} \right.
\end{equation}

In the space 
$\mathcal{X} = \llbracket 1 ; n \rrbracket^{\Nats} \times 
\Bool^n$ we define the distance between 
two points $X = (S,E), Y = (\check{S},\check{E})\in \mathcal{X}$ by%
\begin{eqnarray*}
d(X,Y)& =& d_{e}(E,\check{E})+d_{s}(S,\check{S}), \textrm{ where} \\
\displaystyle{d_{e}(E,\check{E})} & = & \displaystyle{\sum_{k=1}^{n%
}\delta (E_{k},\check{E}_{k})}, \textrm{ and} \\
\displaystyle{d_{s}(S,\check{S})} & = & \displaystyle{\dfrac{9}{n}%
\sum_{k=1}^{\infty }\dfrac{|S^k-\check{S}^k|}{10^{k}}}.%
\end{eqnarray*}

If the floor value $\lfloor d(X,Y)\rfloor $ is equal to $j$,
then the systems $E, \check{E}$ differ in $j$ cells. 
In addition, $d(X,Y) - \lfloor d(X,Y) \rfloor $ is a measure of the differences between strategies $S$ and $\check{S}$. More precisely, this floating part is less than $10^{-k}$ if and only if the first $k$
terms of the two strategies are equal. Moreover, if the $k^{th}$ digit is nonzero, then the $k^{th}$ terms of the two
strategies are different.

In his PhD thesis~\cite{GuyeuxThese10}, Guyeux has already proven that:
\begin{itemize}
\item The function $G_f$ is \emph{continuous} on
 the metric space $(\mathcal{X},d)$. 
\item 
The parallel iterations of $G_\neg$ are \emph{regular}: periodic points of $G_{\neg}$ are dense in $\mathcal{X}$.
\item  The function $G_{\neg}$ is \emph{topologically transitive}: 
for all $X,Y \in
\mathcal{X}$, and for all open balls $B_X$ and $B_Y$ centered in $X$ and
$Y$ respectively, there exist  $X'\in B_X$ and $t \in
\mathds{N}$ such that $G_{\neg}^t(X') \in B_Y$.

\item The function $G_{\neg}$ has 
\emph{sensitive dependence on initial conditions}:
there exists $\delta >0$ such that for any $X\in \mathcal{X}$
and any open ball $B_X$, there exist $X'\in B_X$ and $t\in\Nats$
 such that $d(G_{\neg}^t(X), G_{\neg}^t(X'))>\delta $.

\end{itemize}

To put it differently, a system is sensitive to initial conditions
if any point contains, in any neighborhood, another point with a completely
different future trajectory. Topological transitivity is established when, for any
element, any neighborhood of its future evolution eventually overlaps with any
other open set. On the contrary, a dense set of periodic points is an element of
regularity that a chaotic dynamical system has to exhibit.

We have previously established that the three conditions 
for Devaney's chaos hold for asynchronous iterations.
They thus behave chaotically, as it is defined in the mathematical theory of 
chaos~\cite{devaney,Knudsen94}.
In other words, quoting Devaney in~\cite{devaney}, a chaotic dynamical system ``is unpredictable because of the sensitive dependence on initial conditions. It cannot be broken down or simplified into two subsystems, which do not interact because of topological transitivity. And in the midst of this random behavior, we nevertheless have an element of regularity''.

Intuitively, the topologically transitivity and 
the sensitivity on initial conditions 
respectively address the preimage resistance and the avalanche criteria.
Section~\ref{sec:eval} formalizes this intuition.  

The next section presents our hash function that is 
based on  asynchronous iterations.

\section{Chaos-Based Keyed Hash Function Algorithm}
\label{sec:algo}

The  hash value is obtained as the last configuration 
resulting from iterations of $G_{\neg}$.
We then have to define the pair $X^0=((S^t)^{t \in \Nats},x^0)$, \textit{i.e.},
the strategy $S$ and the initial configuration $x^0$.

\subsection{Computing $x^0$}
\label{subsec:computing x0}
The first step of the algorithm is to transform the message in a normalized
$n = 256$ bits sequence $x^0$.
Notice that this size $n$ of the digest can be changed, 
\emph{mutatis mutandis}, if needed.
Here, this first step is close to the pre-treatment 
of the SHA-1 hash function, but it can easily be replaced by any other compression method.   

To illustrate this step, we take an example, our
original text is: ``\emph{The original text}''.

Each character of this string is replaced by its ASCII code (on 7 bits).
Following the SHA-1 algorithm, 
first we append the character ``1'' to this string, which is then 

\begin{center}
\small
\begin{alltt}
 10101001 10100011 00101010 00001101 11111100
 10110100 11100111 11010011 10111011 00001110
 11000100 00011101 00110010 11111000 11101001.
\end{alltt}
\end{center}

Next we append the block 1111000, which is the binary value of this  
string length (120) and let $R$ be the result.
Finally  another ``1'' is appended to $R$ if and only if 
the resulting length is an even number.

\begin{center}
\small
\begin{alltt}
 10101001 10100011 00101010 00001101 11111100
 10110100 11100111 11010011 10111011 00001110
 11000100 00011101 00110010 11111000 11101001
 1111000.
\end{alltt}
\end{center}

\noindent 
The whole string is copied, but in the opposite direction:

\begin{center}
\small
\begin{alltt}
 10101001 10100011 00101010 00001101 11111100
 10110100 11100111 11010011 10111011 00001110
 11000100 00011101 00110010 11111000 11101001
 11110000 00111110 01011100 01111101 00110010
 11100000 10001101 11000011 01110111 00101111
 10011100 10110100 11111110 11000001 01010011
 00010110 010101.
\end{alltt}
\end{center}

The string whose length is a multiple of 512 is obtained,
by duplicating the string obtained above
a sufficient number of times and truncating it at the next multiple of 512.
This string is further denoted by $D$.
Finally, we split our obtained string into two blocks of 256 bits and apply
to them the exclusive-or (further denoted as XOR) function, from the first two blocks 
to the last one. It results a 256 bits sequence, that is in our example:

\begin{center}
\small
\begin{alltt}
 00001111 00101111 10000010 00111010 00001110
 01100111 01111000 10011101 01010111 00110101
 11010100 01101001 11111001 00011011 01001110
 00110000 11000111 00101101 10001001 11111001
 01100010 10111010 11001110 10101011 10010001
 11101110 01100111 00000101 11000100 00011111
 01001111 00001100.
\end{alltt}
\end{center}
The configuration $x^0$ 
is the result of this pre-treatment  and is a sequence of $n=256$ bits.
Notice that many distinct texts lead to the same string $x^0$. 
The algorithm detailed in~\cite{bcg11:ip} always appends ``1'' to the string
$R$. However such an approach suffered from generating the same 
$x^0$ when $R$'s length is 128. In that case the size of its reverse is again 
128 bits leading a  message of length 256.  
When we duplicate the message, we obtain a message of length 512 composed 
of two equal messages. The resulting XOR function is thus 0 and
this improvement consequently allows us  to avoid this drawback.

Let us build now the strategy $(S^t)^{t \in \Nats}$ that depends on  
the original message and on a given key.

\subsection{Computing $(S^t)^{t \in \Nats}$}
\label{subset:Computing St}

To obtain the strategy $S$, the chaotic proven 
pseudorandom number generator detailed 
in~\cite{DBLP:journals/corr/abs-1112-5239} is used.
The seed of this PRNG is computed as follows: 
first the ASCII code (on 7 bits again) of the 
key is duplicated enough and truncated to the length of $D$.   
A XOR between $D$ and this chain gives the seed of the PRNG, that is left to
generate a finite sequence of natural numbers $S^t$ in 
$\llbracket 1, n \rrbracket$  whose length is $2n$.

\subsection{Computing the digest}
\label{subsec:computing the digest}

To design the digest, asynchronous iterations of $G_{\neg}$ are realized 
with initial state  $X^0=((S^t)^{t \in \Nats},x^0)$ as defined above.
The result of these iterations is a $n=256$ bits vector.
Its components are taken 4 per 4 bits and translated into hexadecimal numbers,
to obtain the hash value:

\begin{center}
\begin{alltt}
  AF71542C90F450F6AE3F649A0784E6B1
  6B788258E87654B4D6353A2172838032.
  \end{alltt}
\end{center}

As a comparison if we replace 
\textquotedblleft \textit{%
  The original text}\textquotedblright\ by \textquotedblleft \textit{the original text}\textquotedblright , the hash function returns:

\begin{center}
\begin{alltt}
  BAD8789AD6924B6460F8E7686A24A422
  8486DC8FDCAE15F1F681B91311426056.
\end{alltt}
\end{center}

We then investigate the qualitative properties of this algorithm.   


\section{Quality Analysis}\label{sec:eval}

We show in this section that, as a consequence of recalled theoretical
results,  this   hash  function  tends  to   verify  desired  informal
properties of a secure keyed one-way hash function.

\subsection{The Strict Avalanche Criterion}
\label{subsec:avalanche}

In our opinion, this criterion is implied
by the  topological properties of sensitive dependence  to the initial
conditions,  expansivity,  and Lyapunov  exponent.  These notions  are
recalled below.

First,  a  function  $f$  has  a  constant  of  expansivity  equal  to
$\varepsilon $ if an arbitrarily  small error on any initial condition
is  \emph{always}  magnified till  $\varepsilon  $.  In our  iteration
context  and  more  formally,  the function  $G_{\neg}$  verifies  the
\emph{expansivity} property if there exists some constant $\varepsilon
>0$ such that for any $X$ and $Y$ in $\mathcal{X}$, $X \neq Y$, we can
find           a            $k\in           \mathbb{N}$           s.t.
$d(G^k_{\neg}(X),G^k_{\neg}(Y))\geqslant \varepsilon$.  We have proven
in~\cite{gfb10:ip}  that,  $(\mathcal{X},G_{\neg})$  is  an  expansive
chaotic system. Its constant of expansivity is equal to 1.


Next,   some  dynamical   systems  are   highly  sensitive   to  small
fluctuations  into   their  initial  conditions.    The  constants  of
sensibility  and   expansivity  have  been   historically  defined  to
illustrate this  fact.  However, in  some cases, these  variations can
become  enormous,  can  grow  in   an  exponential  manner  in  a  few
iterations,  and  neither  sensitivity  nor expansivity  are  able  to
measure such a situation.  This is why Alexander Lyapunov has proposed
a new notion  able to evaluate the  amplification speed of these
fluctuations we now recall:

\begin{definition}[Lyapunov Exponent]
Let be given an iterative system $x^0 \in \mathcal{X}$ and $x^{t+1} = f(x^t)$. 
Its \emph{Lyapunov exponent} is defined by:
$$\displaystyle{\lim_{t \to +\infty} \dfrac{1}{t} \sum_{i=1}^t \ln \left| ~f'\left(x^{i-1}\right)\right|}$$
\end{definition}

By  using  a topological  semi-conjugation  between $\mathcal{X}$  and
$\mathds{R}$, we  have proven in~\cite{GuyeuxThese10}  that, for almost
all $X^0$, the Lyapunov exponent of asynchronous iterations $G_{\neg}$ with
$X^0$ as initial condition is equal to $\ln (n)$.

\bigskip

We can now justify why, in our opinion, the topological properties of the proposed
 hash function lead to the  avalanche effect. Indeed, due to the  sensitive dependence to the
initial condition, two close  media can possibly lead to significantly
different  digests.   The  expansivity  property  implies  that  these
similar medias mostly lead to  very different hash values.  Finally, a
Lyapunov exponent greater than 1 leads to the fact that these two close
media will always end up  by having very different digests.

\subsection{Preimage Resistance}
\label{preimageProof}
\subsubsection{Topological Justifications}

Let us now discuss about the preimage resistance of our keyed hash function denoted by $h$.
As recalled previously, an adversary given a target image $D$ should not be able to find a preimage $M$ such that $h(M)=D$.
One reason (among many) why this property is important is that, on most computer systems, users passwords are stored as the cryptographic hash of the password instead of just the plain-text password.
Thus an attacker who gains access to the password file cannot use it to then gain access to the system, unless it is able to invert target message digest of the hash function.

We now explain why, topologically speaking, our hash function is resistant to preimage attacks.
Let $m$ be the message to hash, $(S,x^0)$ its normalized version (\textit{i.e.}, the initial state of our iteration scheme), and $M=h(m)$ the digest of $m$ by using our method.
So iterations with initial condition $(S,M)$ and iterate function $G_{\neg}$ have $x^0$ as final state.
Thus it is impossible to invert the hash process with a view to obtain the normalized message by using the digest.
Such an attempt is equivalent to try to forecast the future evolution of asynchronous iterations of the $\neg$ function 
by only using a partial knowledge of its initial condition.
Indeed, as $M$ is known but not $S$, the attacker has an uncertainty on the initial condition.
He/she only knows that this value is into an open ball of radius 1 centered at the point $M$, and the number of terms of such a ball is infinite.

With such an incertainty on the initial condition, and due to the numerous chaos properties possessed by our algorithm (as stated in the previous Section), this prediction is impossible.
Furthermore, due to the transitivity property, it is possible to reach all of the normalized medias, when starting to iterate into this open ball.
These qualitative explanations can be formulated more rigorously, by the proofs
given in the next section.

\subsubsection{Proofs of the Second-Preimage Resistance}

We will focus now on a rigorous proof of the second-preimage resistance:
an adversary given a message $m$ should not be able to find another message $m'$ such that $m\neq m'$ and $h(m)=h(m')$.

More 
precisely, we will show that a more general instance of the proposed 
post-treatment described below preserves this
character for a given hash function.

Let 
\begin{itemize}
\item $k_1$,$k_2$,$n$, all in $\mathds{N}^*$, where 
$k_1$ is the size of the key,
$k_2$ is the size of the seed,
and $n$ is the size of the hash value,
\item $h:(k,m) \in \mathds{B}^{k_1}\times\mathds{B}^* \longmapsto h(k,m) \in \mathds{B}^n$ a keyed hash function,
\item $S:k\in \mathds{B}^{k_2} \longmapsto S(k) \in  \llbracket 1,n\rrbracket$ a cryptographically secure pseudorandom number generator, 
\item $\mathcal{K}=\mathds{B}^{k_1}\times\mathds{B}^{k_2}\times \mathds{N}$ called the \emph{key space}, 
\item and $f:\mathds{B}^n \longrightarrow \mathds{B}^n$ a bijective map.
\end{itemize}

We define the keyed hash function $\mathcal{H}:\mathcal{K}\times\mathds{B}^* \longrightarrow \mathds{B}^n$ by the following procedure\\

\begin{tabular}{ll}
\underline{\textbf{Inputs:}} & $K = (K_1,K_2,N)\in \mathcal{K}$\\
                            & $m \in \mathds{B}^*$\\
\underline{\textbf{Runs:}} & $X=h(K_1,m)$\\
                           & for $i=1, \hdots, N:$\\
                           & ~~~~ $X=G_f(S^i(K_2),X)$\\
                           & return $X$
\end{tabular}

\noindent where $K_1$ is the key of the inputted 
hash function, $K_2$ the seed of the strategy 
used in the post-treatment iterations,
where $N$ is for the size of
this strategy. We have the following result.

\begin{theorem}\label{th:pr}
If $h$ satisfies the second-preimage resistance property, 
then it is the case for $\mathcal{H}$ too.
\end{theorem}

To achieve the proof, we introduce the two following lemmas.

\begin{lemma}
\label{lemma1}
If $f: \mathds{B}^n \longrightarrow \mathds{B}^n$ is bijective,
then for any $S \in \llbracket 1,n \rrbracket$, 
the map 
$G_{f,S}: x \in \mathds{B}^n \rightarrow G_f([S,1,\ldots,1],x)_2 \in \mathds{B}^n$ is bijective too where $G_f(\_,\_)_2$ is the 
second term of the pair $G_f(\_,\_)$.
\end{lemma}
\begin{IEEEproof}
  Since $\mathds{B}^n$ is a finite set, it is sufficient to prove 
  that $G_{f,S}$ is surjective.
  Let $y=(y_1, \hdots, y_n) \in \mathds{B}^n$ and 
  $S\in \llbracket 1,n \rrbracket$.
Thus 
$G_{f,S}\left((y_1,\hdots,y_{S-1},f^{-1}(y_S),y_{S+1},\hdots, y_n)\right)_2=
G_f\left([S,1,\ldots,1],
  (y_1,\hdots,y_{S-1},f^{-1}(y_S),y_{S+1},\hdots, y_n)\right)_2=
\left([1,\ldots,1],
(y_1,\hdots,y_{S-1},y_S,y_{S+1},\hdots, y_n)\right)_2=
(y_1,\hdots,y_{S-1},y_S,y_{S+1},\hdots, y_n)=
y$.
So $G_{f,S}$ is a surjective map between two finite sets and thus bijective.
\end{IEEEproof}

\begin{lemma}
\label{lemma2}
Let $S \in \llbracket 1,n \rrbracket^\mathds{N}$
and $N \in \mathds{N}^*$. If $f$ is bijective, then 
$G_{f,S,N}:x \in \mathds{B}^n \longmapsto G_f^N(S,x)_2 \in \mathds{B}^n$
is bijective too.
\end{lemma}

\begin{IEEEproof}
Indeed, $G_{s,f,n}=G_{f,S^n}\circ \hdots \circ G_{f,S^0}$ is bijective as a composition 
of bijective maps (as stated in Lemma~\ref{lemma1}).
\end{IEEEproof}

We are now able to prove theorem~\ref{th:pr}.

\begin{IEEEproof}
Let $m,k \in \mathds{B}^*\times\mathcal{K}$. If a message $m'\in\mathds{B}^*$ can be found such that $\mathcal{H}(k,m)=\mathcal{H}(k,m')$, then, according to Lemma~\ref{lemma2}, $h(k_1,m)=h(k_1,m')$: a second-preimage for $h$ has thus been
found.
\end{IEEEproof}

\subsection{Algorithm Complexity}

In this section the complexity of the above hash  function is evaluated  
for a size $l$ of the media (in bits).

\begin{theorem}\label{th:cplx}
Let $l$ be the size of the message to hash and $n$
be the size of its hash value. 
The algorithm detailed along these lines requires 
$\mathcal{O}(l) + \mathcal{O}(n^2)$ elementary operations
to produce the hash value.
\end{theorem}

\begin{IEEEproof}
In the $x^0$ computation stage only linear operations over
binary tables are achieved. More precisely it first executes  
one ASCII translation yielding a message of length $7l$, 
a length computation that increases the message length of 
$\log_2(7l)$. One bit is possibly added.
Thus a reversed copy that leads to a message of length $l'$ 
that is $14l + 2+ 2\log_2(7l)$. 
The number of duplication steps to get a message whose length is greater than
a multiple of $2n$ is formally given by
\begin{equation}\label{eq:min}
\min_{k \ge 1} \{k \mid \exists p \,.\, p \ge 1 \land  k \times l' \ge 2np \}
\end{equation}

This number is bounded by 
$$ 
k'=\max \{1,n \}.
$$
If $14l + 2 + 2\log_2(7l)$ is greater than $2n$ it is sufficient to 
duplicate the message once. 
Otherwise, $\lfloor 1+ \dfrac{2n}{14l+ 2 + 2\log_2(7l)}\rfloor$ is greater than
$\dfrac{2n}{14l+ 2 + 2\log_2(7l)}$ and thus 
$l' \times \dfrac{2n}{14l+ 2 + 2\log_2(7l)}$
is greater than  $2n$ and there exists a $p$ ($p=1$) such that 
$k\times l' \ge 2np$.
Thus the minimum of the set given in Eq.(\ref{eq:min}) is less 
than $\lfloor 1+ \dfrac{2n}{14l+ 2 + 2\log_2(7l)}\rfloor$, 
which is less than $n$.

To sum up, the initialization of $x^0$  
requires at most $k'+l'$ elementary operations.

Let us now detail the $S$ computation step. The number of 
elementary operations to provide the seed is bounded by 
$k'+l'$.
Next, the embedded PRNG~\cite{DBLP:journals/corr/abs-1112-5239}, 
that combines the XORShift, xor128, and XORWow PRNGs  
requires 35 elementary operations 
(17 XOR, 13 rotations, and 5 arithmetic operations) 
for generating a 32 bits number and thus $35\dfrac{2n}{32}$ to 
get a number on $2n$ bits. 
Furthermore, since the strategy length is $2n$, 
the computation of $S$ requires at most $k'+l'+ 2n \times 35\dfrac{2n}{32}$,
which is less than $k' + l' + 5n^2$.

At least, since each iteration modifies 
only one bit, iterations require $2n$ elementary operations.

   
Finally, we have at most $2k' + 2l' + 5n^2 + 2n$  elementary operations to 
provide a hash value of size $n$. This bound is in 
$\mathcal{O}(l + n^2)$.  
\end{IEEEproof}



\section{Experimental Evaluations}

Let us now give some examples of hash values
before statistically studying the quality of hash outputs.


\subsection{Examples of Hash Values}\label{sub:exh}

\begin{figure}
\centering
\includegraphics[scale=0.25]{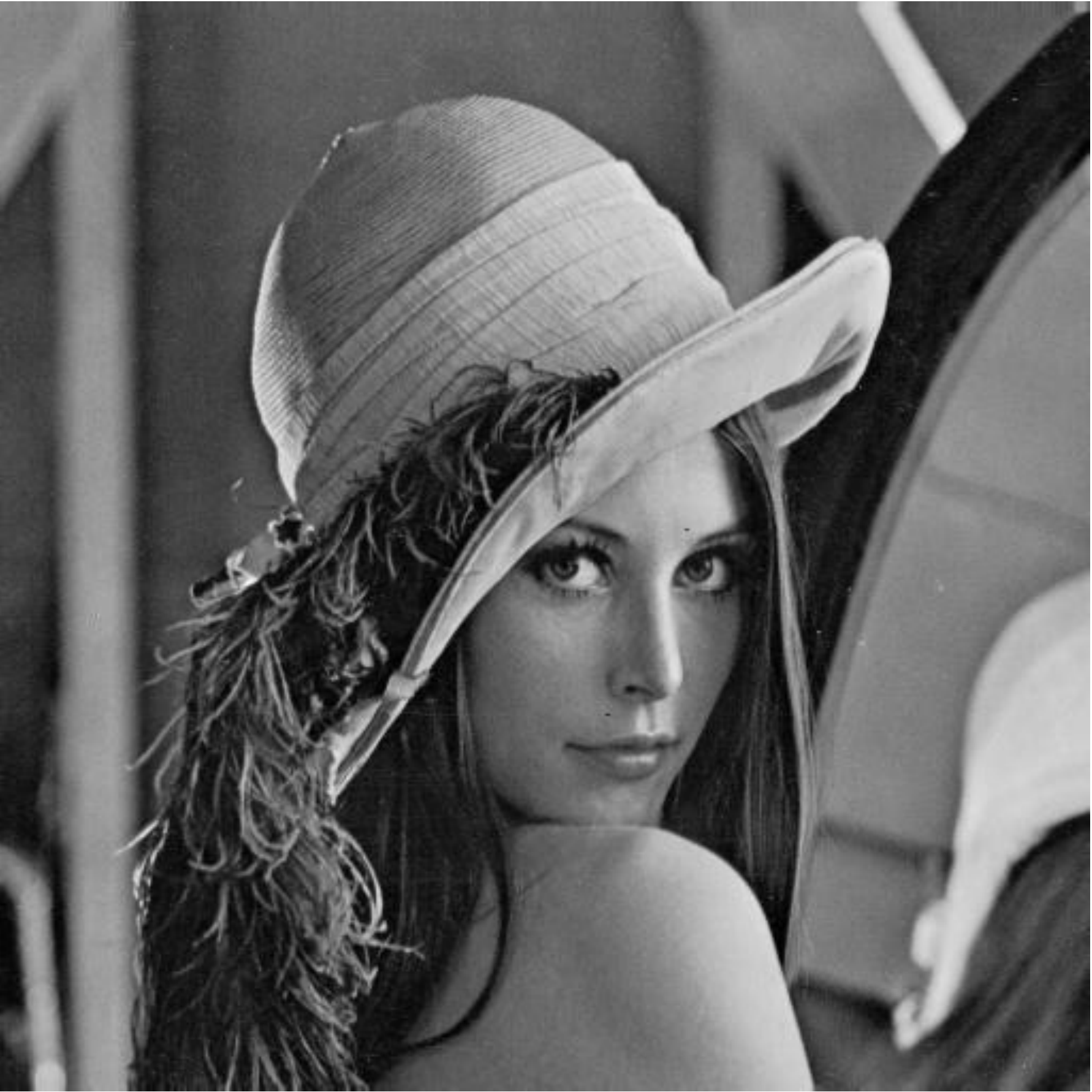}
\caption{The original plain-image.}
\label{plain-image}
\end{figure}

Let  us    consider  the proposed  hash  function  with  $n=256$.
We consider the key to be equal to ``\textit{my key}''.
To  illustrate  the confusion and  diffusion properties~\cite{Shannon49},
we  use this function to generate hash values in the following cases:
\begin{description}
\item[Case 1.]~ The original text message is the poem \textit{Ulalume}
(E.A.Poe),  which is  made of  104 lines,  667 words,  and $3,754$
characters.
\item[Case 2.]~ We change  \textit{serious} by \textit{nervous} in the
verse ``\textit{Our talk had been serious and sober}''
\item[Case 3.]~ We replace the last point `.' with a coma `,'.
\item[Case 4.]~  In ``\textit{The skies they were  ashen and sober}'',
skies becomes Skies.
\item[Case  5.]~ The  new original  text is  the binary  value  of
Figure~\ref{plain-image}.
\item[Case 6.]~  We add 1  to the gray  value of the pixel  located in
position (123,27).
\item[Case 7.]~ We subtract 1 to  the gray value of the pixel located
in position (23,127).
\end{description}

The corresponding hash values in hexadecimal format are:

\begin{description}
\item[Case 1.]~\textsf{
    0B4730459FBB5E54A18A9CCD676C8396
    365B0104407D98C866FDAA51A07F0E45},
\item[Case 2.]~\textsf{
    752E28088150B98166D870BC24177342
    23A59463D44B83E9808383B30F8B8409},
\item[Case 3.]~\textsf{
    C10EED0A9D44856847F533E5647D0CCD
    2C58A08643E4D3E5D8FEA0DA0E856760},
\item[Case 4.]~\textsf{
    52BF23429EC3AD16A0C9DE03DF51C420
    4466285448D6D73DDFB42E7A839BEE80},
\item[Case 5.]~\textsf{
    5C639A55E2B26861EB9D8EADDF92F935
    5B6214ADC01197510586745D47C888B8},
\item[Case 6.]~\textsf{
    E48989D48209143BAE306AC0563FFE31
    EAB02E5E557B49E3442A840996BECFC1},
\item[Case 7.]~\textsf{
    EC850438A2D8EA95E691C746D487A755
    12BEE63F4DDB4466C11CD859671DFBEB},
\end{description}

 These simulation results are coherent with the topological properties
of  sensitive dependence  to the  initial condition,  expansivity, and
Lyapunov exponent: any alteration  in the message causes a substantial
difference in the final hash value.

\subsection{Statistical Evaluation of the Algorithm}

We  focus now  on  statistical studies of  diffusion and  confusion
properties.  Let  us recall that  confusion refers to
the desire to make the relationship between the key and the digest
as complex and involved as  possible, whereas diffusion means that the
redundancy in the statistics of  the plain-text must be "dissipated" in
the statistics of the  cipher-text.  Indeed, the avalanche criterion is
a modern  form of the  diffusion, as this  term means that  the output
bits should depend on the input bits in a very complex way.

\begin{figure}[t]
\centering
\subfloat[Original text]{
\includegraphics[scale=0.45]{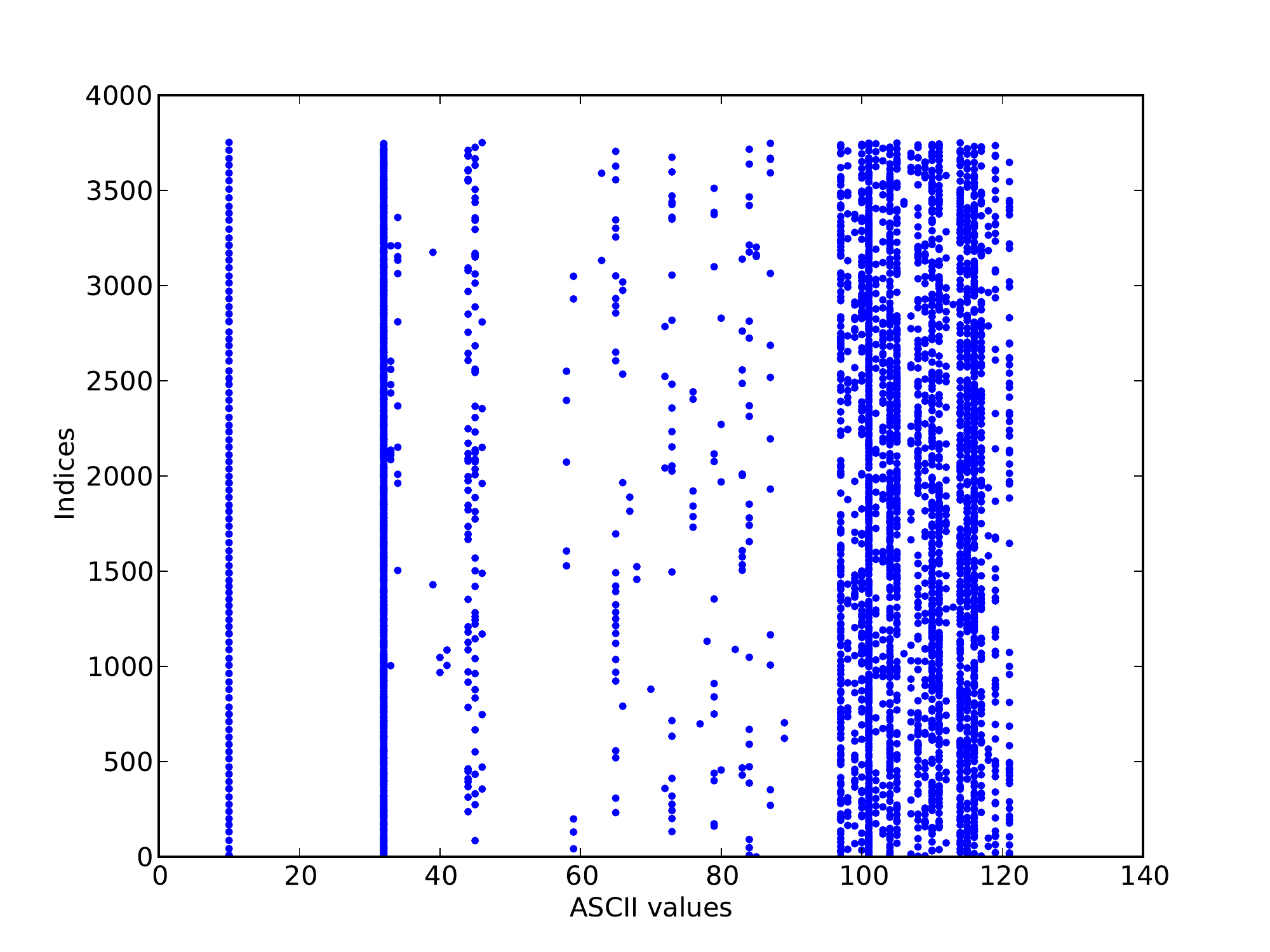}
\label{fig:ASCII repartition0}
} 

\subfloat[Digest]{
\includegraphics[scale=0.45]{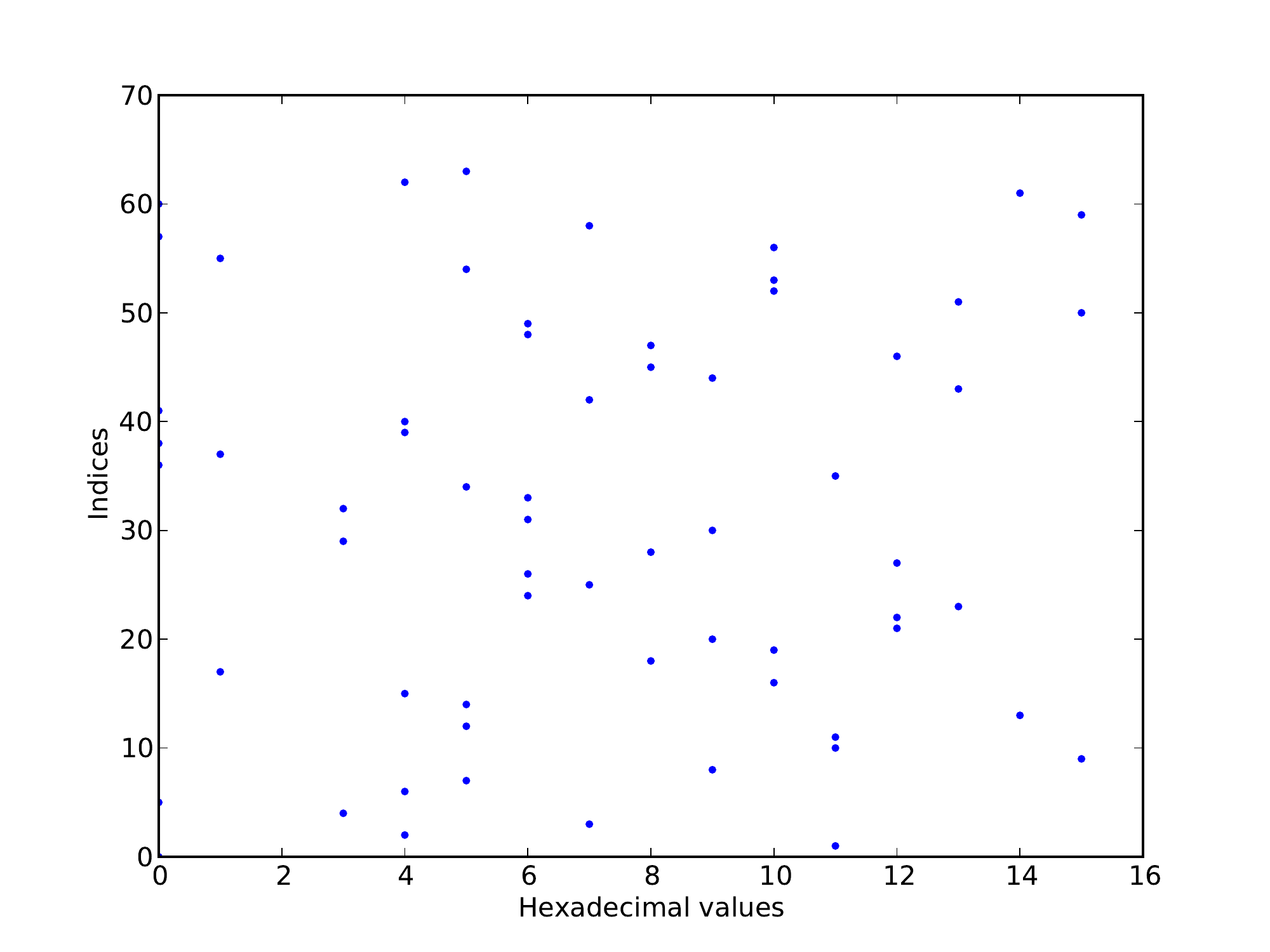}
\label{fig:Hexa repartition0}
}
\caption{Values repartition of Ulalume poem}
\end{figure}

\subsubsection{Uniform repartition for hash values}

To show the diffusion and confusion properties verified by our scheme,
we first  give  an  illustration  of  the  difference  of  characters
repartition between a plain-text and  its hash value, when the original
message is again the Ulalume poem.
In Figure~\ref{fig:ASCII  repartition0}, 
(resp. in Figure~\ref{fig:Hexa repartition0})
the X-axis represents ASCII numbers 
(resp. hexadecimal numbers) whereas 
the Y-axis gives for each X-value its position in the original text
(resp. in the digest). 
For instance, in Figure~\ref{fig:Hexa repartition0}, the point 
$(1,17)$ means that the character 1 is present in the digest at position 
$17$ (see Case 1, Section.~\ref{sub:exh}).
We can see that ASCII codes are localized within a small area
(e.g., the ASCII ``space'' code and the lowercase characters), 
whereas in Figure~\ref{fig:Hexa repartition0}  the hexadecimal numbers of 
the hash value are uniformly distributed.

\begin{figure}[t]
\centering
\subfloat[Original text]{
\includegraphics[scale=0.45]{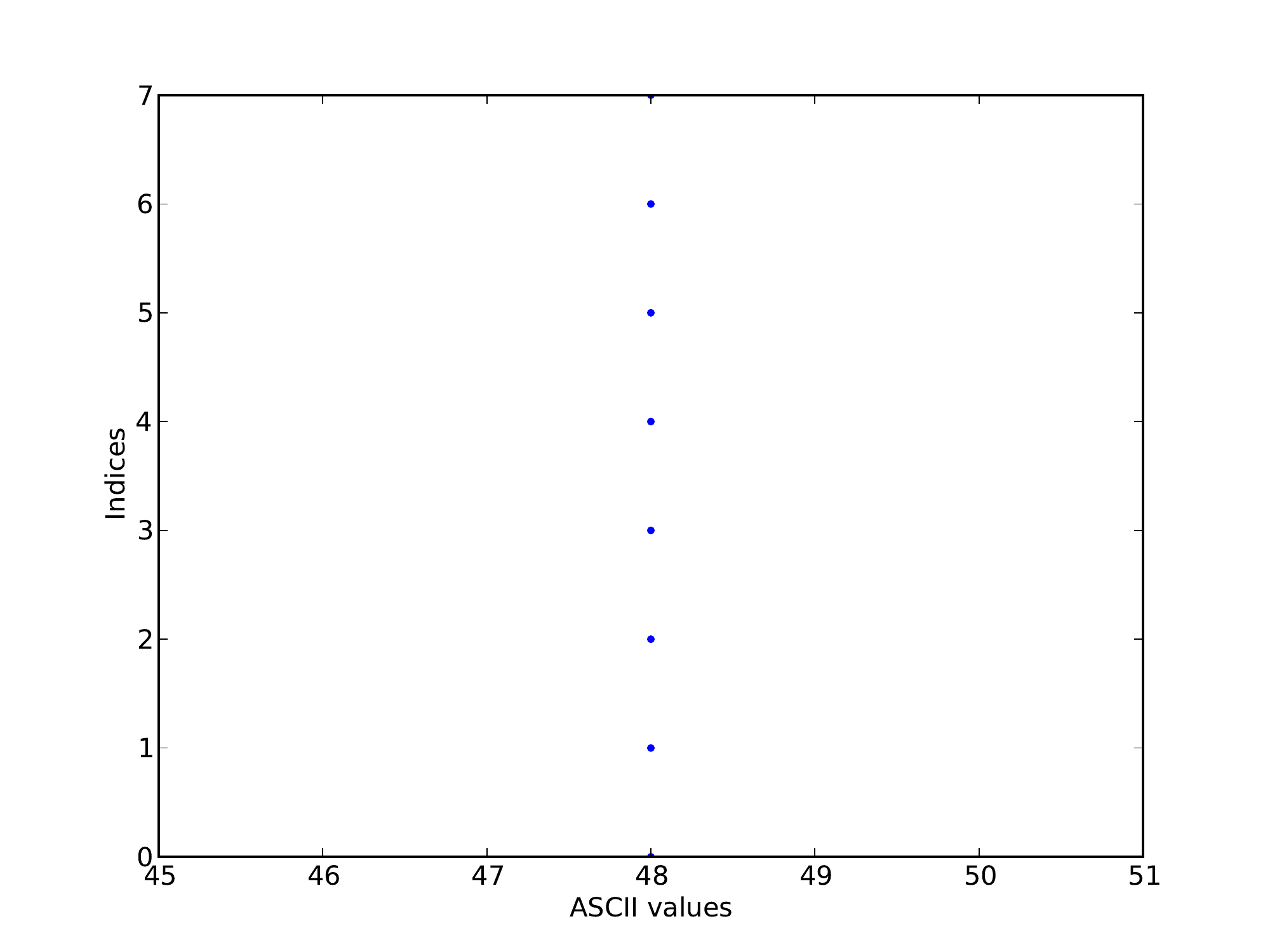}
\label{fig:ASCII repartition}
}

\subfloat[Digest]{
\includegraphics[scale=0.45]{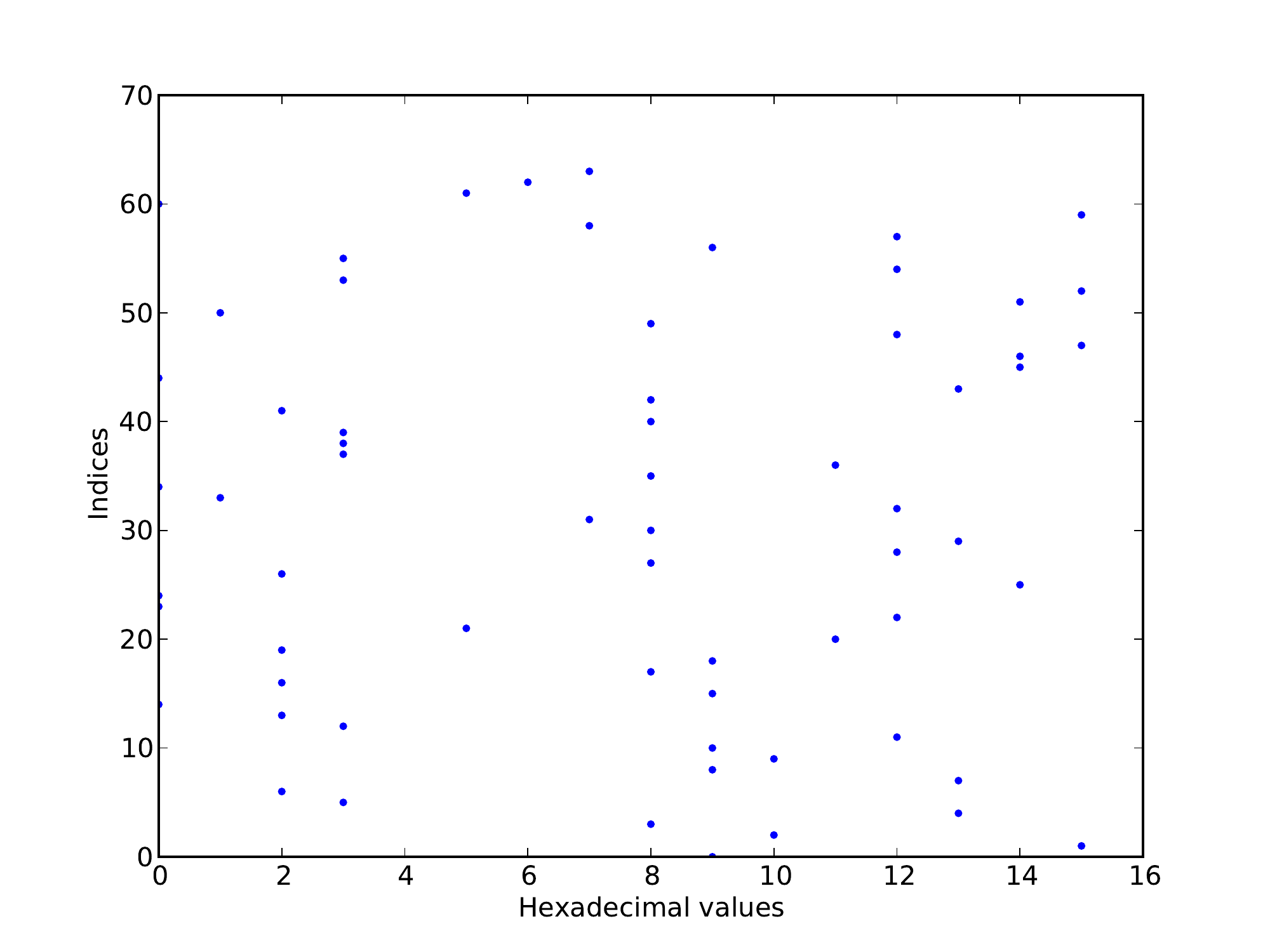}
\label{fig:Hexa repartition}
}
\caption{Values repartition of the ``\textit{00000000}'' message}
\end{figure}

A similar experiment has been realized with a message having the same size,
but which is only constituted by the character ``\textit{0}''. 
The contrasts between the plain-text message and its digest 
are respectively presented in Figures~\ref{fig:ASCII repartition} 
and~\ref{fig:Hexa repartition}.
Even under this very extreme condition, the distribution of the digest still remains uniform. 
To conclude, these simulations tend to indicate that no information concerning the original message can be found into its hash value, as it is recommended by the Shannon's diffusion and confusion requirements.

\subsubsection{Behavior through small random changes}

\begin{figure}
\centering
\includegraphics[scale=0.56]{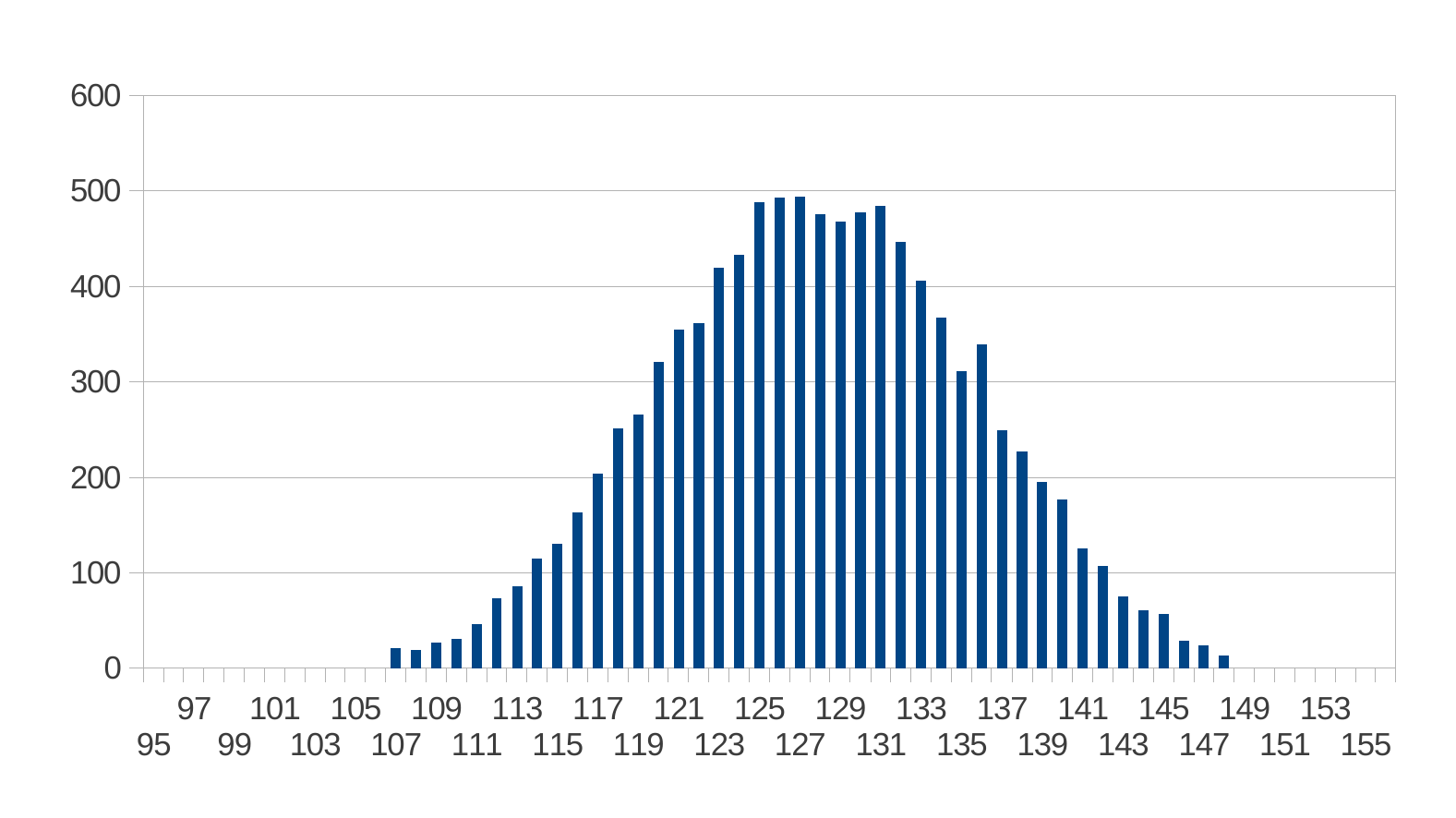}
\caption{Histogram}
\label{Histogram}
\end{figure}

We now consider the following experiment. A first message of 1000 bits is
randomly generated, and its hash value of size $n=256$ bits is computed.
Then one bit is randomly toggled into this message and the digest of
the new message is obtained.
These two hash values are compared by using the hamming distance, 
to compute the number $B_i$ of changed bits.
This test is reproduced $t=10,000$ times.
The corresponding distribution of $B_i$ is presented in Figure~\ref{Histogram}.

As desired, Figure~\ref{Histogram} shows that the distribution
is centered around 128, which reinforces the confidence put
into the good capabilities of diffusion and confusion of the proposed hash algorithm.
To analyse these results, the following common statistics are used.
\begin{itemize}
\item Mean changed bit number 
  $$\overline{B} = \frac{1}{t}\sum_{i=1}^{t} B_i.$$
\item Mean changed probability 
  $$P = \frac{\overline{B}}{n}.$$
\item 
  $\Delta B = \sqrt{\dfrac{1}{t}
    \sum_{i=1}^{t} (B_i-\overline{B})^2}$.
\item 
  $\Delta P = \sqrt{\dfrac{1}{t}\sum_{i=1}^{t} (\frac{B_i}{n}-P)^2}$.
\end{itemize}

The obtained statistics are listed in 
Table~\ref{table:statistical performances} where $n$ belongs to $\{256;512;1,024\}$.
In that study, starting from a message of length $1,000$ and its digest, 
all the messages that have one bit of difference are further generated 
and the  digest of the new message is obtained.
Obviously, both the mean changed bit number 
$\overline{B}$ and the mean changed probability $P$ are close to the ideal
values ($\frac{n}{2}$ bits and 50\%, respectively), which illustrates the diffusion and confusion capability of our algorithm. 
Lastly, as $\Delta B$ and $\Delta P$ are very small, 
these capabilities are very stable.

\setlength{\extrarowheight}{5 pt}
\begin{table}
\begin{tabular}{ccccccc}
\hline
 & $B_{min}$ & $B_{max}$ & $\overline{B}$ & $P(\%)$ & $\Delta B$ & $\Delta P(\%)$ \\

\hline
$n = 256$  & 87 & 167 & 127.95  & 49.98  & 8.00 & 3.13  \\
$n = 512$  & 213 & 306 & 255.82 & 49.97 & 11.29 & 2.21  \\
$n = 1024$ & 446 & 571 & 511.54 & 49.96 & 15.97 & 1.56  \\
\hline
\end{tabular}
\caption{Statistical performances of the proposed hash function}
\label{table:statistical performances}
\end{table}
\enlargethispage{-0.4in}
\setlength{\extrarowheight}{0 pt}




\subsection{Strict Avalanche Criterion Evaluation}\label{sub:sac}
This section focuses on checking whether the developed hash function 
verifies the strict avalanche criterion, as given in Definition~\ref{def:sac}.
Quoting remarks of~\cite{Webster:1985:DS:646751.704578}, ``\emph{Unless $n$
is small, it would be an immense task to follow this procedure for all 
possible vector pairs $x$ and $\overline{x}^i$}''.
The authors propose thus the alternative method of computing a dependence matrix
$J$  of size $m\times n$ 
between the $j$-th, $1 \le j \le m$, element of the digest and  
$i$-th, $1 \le i \le n$, element of the original message. 
A simulation consists in first randomly choosing the size $n$ of the
message to hash (100 values in $\llbracket 1,1000 \rrbracket$ for us).
Next, a set of large size $r$ ($r=1,000$ in our case) 
of messages $x$
is randomly computed. For each of them, the set 
$\{\overline{x}^1,\ldots,\overline{x}^n\}$ is formed such that 
$x$ and $\overline{x}^i$ only differ in bit $i$. 
The set of $m$-bit vectors 
$$
\{
f(x) \oplus f(\overline{x}^1),
\ldots,
f(x) \oplus f(\overline{x}^n) \}$$
is thus  computed where $f$ is the hash function applied on vector of bits.
The value of bit $i$ (either a 1 or a 0) in  
$\big(f(x) \oplus f(\overline{x}^i)\big)_j $ is added to $J_{ij}$.
Finally each element of $J$ is divided by $r$. 
If every $J_{ij}$ are close to one half, the strict avalanche criterion 
is established. 
For all these experiments, the average value of $J_{ij}$ is 0.5002,
the minimal value is 0.418, the maximal value is 0.585, and the standard 
deviation is 0.016.

\section{Conclusion}
In this research work, the hash function proposed in the Third International Conference on Evolving Internet, INTERNET11
(June 19-24, 2011, Luxembourg)~\cite{bcg11:ip} has been completely rethought. 
The second-preimage resistance
has been proven, leading to better experimental results for the proposed hash function.
Moreover,
we have shown that
this function 
has a complexity 
that can be expressed as a polynomial function of the message length and 
of the digest size.    
Finally, we have statistically established that our function verifies the SAC.

If we now consider our approach as an asynchronous iterations post-treatment of 
an existing hash function. The security of this hash function is reinforced
by the unpredictability of the behavior of the proposed post-treatment.
Thus, the resulting hash function, a combination between an existing hash function and asynchronous iterations,   satisfies important properties of topological chaos such as sensitivity to initial conditions, uniform repartition (as a result of the transitivity), unpredictability, and expansivity.
 Moreover, its Lyapunov exponent can be as great as needed.
The results expected in our study have been experimentally checked. The choices made in this first study are simple: initial conditions designed by using the same ingredients as in the SHA-1, negation function for the iteration function, \emph{etc.} 
But these simple choices have led to desired results, justifying that such a post-treatment can
possibly improve the security of the inputted hash function.
And, thus, such an approach should be investigated more largely.

This is why, in future work, we will test other choices of iteration functions and 
strategies. We will try to characterize topologically the diffusion and confusion capabilities.  Other properties induced by topological chaos will be explored and their interest for the realization of hash functions will be deepened.
Furthermore, other security properties of resistance and pseudo-randomness will be proven.
We will thus compare the results of this post-treatment on several hash functions, among other things with the SHA-3 finalists~\cite{nistSha3comp}.





\section*{Acknowledgment}
The authors are grateful to the anonymous 
reviewers for their suggestions to improve the quality of the paper.

\bibliographystyle{plain}
\bibliography{mabase}

\end{document}